# The Structure of CaSO$_4$ Nanorods - the Precursor of Gypsum

*Tomasz M. Stawski[1]\*, Alexander E.S. Van Driessche[2]\*\*, Rogier Besselink[1,2],*

*Emily H. Byrne[3], Paolo Raiteri[3]\*\*\*, Julian D. Gale[3], Liane G. Benning[1,4,5]*

[1]*German Research Centre for Geosciences, GFZ, Interface Geochemistry, Telegrafenberg, 14473, Potsdam, Germany;*

[2]*Université Grenoble Alpes, Université Savoie Mont Blanc, CNRS, IRD, IFSTTAR, ISTerre, 38000, Grenoble, France;*

[3]*Curtin Institute for Computation, Curtin University, PO Box U1987, Perth, WA 6845, Australia*

[4]*Department of Earth Sciences, Free University of Berlin, Malteserstr. 74-100 / Building A, 12249, Berlin, Germany;*

[5]*School of Earth and Environment, University of Leeds, Woodhouse Lane, LS2 9JT, Leeds, UK.*

*Corresponding authors:*

*\* tomasz.stawski@gmail.com; www.researchgate.net/profile/Tomasz_Stawski*

*\*\* alexander.van-driessche@univ-grenoble-alpes.fr*

*\*\*\* P.Raiteri@curtin.edu.au*

**Abstract**

Understanding the gypsum ($CaSO_4 \cdot 2H_2O$) formation pathway from aqueous solution has been the subject of intensive research in the past years. This interest stems from the fact that gypsum appears to fall into a broader category of crystalline materials whose formation does not follow classical nucleation and growth theories. The pathways involve transitory precursor cluster species, yet the actual structural properties of such clusters are not very well understood. Here, we show how in situ high-energy X-ray diffraction experiments and molecular dynamics (MD) simulations can be combined to derive the structure of small $CaSO_4$ clusters, which are precursors of crystalline gypsum. We fitted several plausible structures to the derived pair distribution functions and explored their dynamic properties using unbiased MD simulations based on both rigid ion and polarizable force fields. Determination of the structure and (meta)stability of the primary species is important from both a fundamental and applied perspective; for example, this will allow for an improved design of additives for greater control of the nucleation pathway.

**Introduction**

In recent years, we have come to appreciate the astounding complexity of the processes leading to the formation of mineral phases from ions in aqueous solutions. The original, and rather naive, 'textbook' image of these phenomena, stemming from the adaptation of classical nucleation and growth theories, has increased in complexity due to the discovery of a variety of precursor and intermediate species. These include solute clusters (e.g. prenucleation clusters, PNCs), liquid(-like) phases, as well as amorphous and nanocrystalline solids, among others[1–9]. Such precursor or intermediate species constitute different points along the pathway from dissolved ions to the final solids (e.g. crystals)[10]. Despite a plethora of indirect experimental[4] and modeling-based[11] evidence for the existence of such precursor species there has yet to be any direct structural information derived from experimental observations. Recently, we reported[12,13] that elongated sub-3 nm primary species are the fundamental building units for dihydrated calcium sulfate crystals, i.e. gypsum ($CaSO_4 \cdot 2H_2O$). Based on in situ and time-resolved small- and wide-angle X-ray scattering data (SAXS/WAXS), we showed that the formation, aggregation, rearrangement and the partial coalescence of these structural units constitutes the actual mechanism behind gypsum crystallization. Nevertheless, from those experiments we were unable to elucidate the nature or the molecular structure of the primary species. To address this issue, we carried out in situ high-energy X-ray diffraction experiments and combined these with molecular dynamics simulations. From both in situ and in silico data we have derived the molecular structure of the precursor species. Determination of the actual structure and (meta)stability of the primary species is important both from a fundamental and applied perspective; for example, this will allow for improved design of additives for greater control of the nucleation pathway.

**Experimental**

*Synthesis and scattering experiments*

$CaSO_4$ was synthesized by reacting equimolar aqueous solutions (ultrapure deionized water, MilliQ, resistivity >18 MΩ·cm) of $CaCl_2 \cdot 2H_2O$ (pure, Sigma) and $Na_2SO_4$ (> 99%, Sigma) at final concentrations of 50 and 100 mmol/L at T = 10 °C. Prior to mixing, all solutions were equilibrated in a fridge at the desired reaction temperatures. The compositions of solutions (Table 1), under the assumption that gypsum was the final solid phase, were calculated with PHREEQC[14] based on the following reaction:

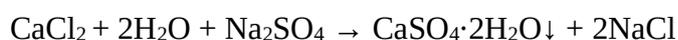

$CaCl_2 + 2H_2O + Na_2SO_4 \rightarrow CaSO_4 \cdot 2H_2O\downarrow + 2NaCl$

$CaSO_4$ formation reactions were performed in a 200 mL temperature-stabilized glass reactor. The reaction temperature in the reactor was maintained by means of an active cooling circulator (Huber Ministat 125), and the feedback thermocouple was placed in the reactor. The reacting solutions were continuously stirred at 350 rpm, and circulated through a custom-built PEEK flow-through cell with Kapton capillary (ID 1.5 mm) using a peristaltic pump (Gilson MiniPuls 3, flow ~10 mL/second). Typically, an experiment started with 40 mL of a temperature-stabilized $CaCl_2$ aqueous solution inside the reactor. This solution was circulated through the capillary cell while scattering patterns were collected continuously as described below. $CaSO_4$ formation reactions were initiated remotely through the injection of 40 mL of a temperature-stabilized Na$_2$SO$_4$ aqueous solution. Injection and mixing (within 30 s) of the two solutions was achieved with the use of a secondary peristaltic pump (Gilson MiniPuls 3) as a remote fast injection system. Depending on the supersaturation, reactions were followed for up to 4 hours. The experimental approach used for these in situ HEXD experiments is identical to that used in our previous in situ SAXS experiments.

**Table 1.** Composition of CaSO$_4$ solutions from PHREEQC[14] modelling.

|  | **50 mM CaSO$_4$** | **100 mM CaSO$_4$** |
|---|---|---|
| **Initial composition** | [Ca$^{2+}$] = [SO$_4^{2-}$] = 35 mM<br>[CaSO$_4$]* = 15 mM | [Ca$^{2+}$] = [SO$_4^{2-}$] = 69 mM<br>[CaSO$_4$]* = 31 mM |
| **Final composition** | [Ca$^{2+}$] = [SO$_4^{2-}$] = 16.3 mM<br>[CaSO$_4$]* = 3.9 mM<br>[solid equivalent] = 29.8 mM | [Ca$^{2+}$] = [SO$_4^{2-}$] = 20 mM<br>[CaSO$_4$]* = 3.9 mM<br>[solid equivalent] = 76.1 mM |

\* - dissolved species. The model considers here an ion pair.

In order to follow the formation processes in situ, and in a time-resolved manner, the flow-through cell was mounted on a high-energy X-ray diffraction (HEXD) system and the measurements were carried out at beamline I15 of the Diamond Light Source (UK). The atomic pair distribution functions (PDFs) were obtained by using the PDFgetX3 software package[15]. Details regarding the data collection and analysis are provided in the ESI.

*Atomistic simulations*

Classical molecular dynamics simulations were run with the LAMMPS program[16] at 300 K and 1 atm in a cubic simulation cell of approximately 50 Å in side, which contained ~4,130 water molecules and the relevant $CaSO_4$ cluster. The long-range electrostatics were computed with the Particle Mesh Ewald algorithm with an accuracy of $10^{-5}$ while the bonded and van der Waals interactions were described by our recently developed force field[17] (FF1). The real space cut-off was set to 9 Å with a tapering function applied between 6 and 9 Å. The water was described with the SPC/Fw model[18], and no tapering was applied to the water-water interactions. The equations of motions were integrated with a 1 fs time-step using the velocity Verlet algorithm, while the temperature was controlled with the Canonical Sampling through Velocity Rescaling algorithm[19]. The cell was constrained to remain cubic throughout the simulations.

All simulations with the AMOEBA polarizable force field[20] were run with the openMM7.1 code[21]. The equations of motion were integrated with the Langevin algorithm using a 0.5 fs time-step, a friction coefficient of 1 $ps^{-1}$ and a temperature of 300 K. The pressure was maintained at 1 atm using an isotropic Monte Carlo barostat[22,23] with a collision frequency of 25 steps. The real space cut-off was set to 10 Å for all non-bonded interactions, while the long-range electrostatic interactions were computed with the Particle Mesh Ewald algorithm, again with an accuracy of $10^{-5}$. Induced dipoles were solved for iteratively at each time-step until an accuracy of $10^{-5}$ was reached. The initial configurations were the same as used in the rigid-ion MD simulations.

The trajectories for cluster I calculated for both force fields are included as a part of the ESI: Supporting Files.

**Results and Discussion**

Background-corrected and time-resolved in situ diffraction patterns from a 50 mM $CaSO_4$ solution at 10 °C are presented in Fig 1A. From the very first frame (265 s) and during the initial ~2,000 s, the measured scattering data contains only an invariant broad diffuse pattern i.e. diffraction peaks were absent. After this induction time, the diffraction pattern of a crystalline phase gradually started to develop. The crystalline phase coexisted throughout the duration of the measurements with the diffuse pattern (Fig. 1B). Importantly, the as-formed diffuse pattern is significantly different from the scattering patterns obtained from the solvent, i.e. $H_2O$, and the aqueous solutions containing the various components of the system under consideration (Fig. 1C). Therefore, the diffuse pattern can be attributed to the presence of a disordered phase of $CaSO_4$ as is demonstrated below. Selected diffraction patterns were converted to pair distribution functions (PDFs, $G(r)$s) using a numerical inverse Fourier transform, in order to analyse, identify and model the structural changes in real-space. Throughout the reaction, the crystalline material was fitted with a gypsum model structure (Fig. 1D) without any trace of other crystalline $CaSO_4$ phases (i.e. bassanite or anhydrite). The $G(r)$ resulting from the diffuse scattering shows several inter-atomic distances common to all calcium sulfate phases (Fig. 1E): S-O at 1.49 Å; Ca-O at ~2.6 Å; shortest (1$^{st}$ neighbour) Ca-Ca and S-S at ~4 Å. With respect to the crystalline forms this disordered phase lacks the longer, i.e. 2nd neighbour, Ca-Ca/Ca-S distance found at ~5-6 Å, although other longer distances are still present. Finally, and most importantly, the PDF plot shows that the structural coherence of the species constituting the diffuse phase does not exceed ~1.5 nm, and thus we can refer to it as a nanophase. Overall, the PDF plots reveal that the species from which the nanophase is built are structurally more complex than simple ions or ion pairs in solution. They have the form of clusters containing at least one single Ca-Ca distance.

The diffuse diffraction patterns shown in Fig. 1A-C are lacking detail, and only show very broad diffraction maxima, which is characteristic of small and thermally labile species in solution. Consequently, one expects that for the primary species a whole family of possible structures can fit to the $G(r)$, rather than one unique solution. Based on our previous small-angle X-ray scattering results, we can narrow down the number of possibilities by constraining the form of such clusters to rod-shaped species with a cross-sectional diameter of ~0.5 nm and a length of several nm[12,13]. Taking into account these constraints, we constructed template clusters, whose structure was further optimized to fit the experimental $G(r)$. Our template clusters were constructed by choosing a primary motif, which contained at least one single Ca-Ca bond distance, stacking a number of such motifs to obtain a rod, and then optimizing the structure using the randomized simplex method so that the simulated PDF of the structure would match as closely as possible the experimental one. In Fig. 2 we present several possible structures of the primary species, based on the best fits to the $G(r)$ of the nanophase obtained by optimizing the initial template clusters using a DiffPy library[24].

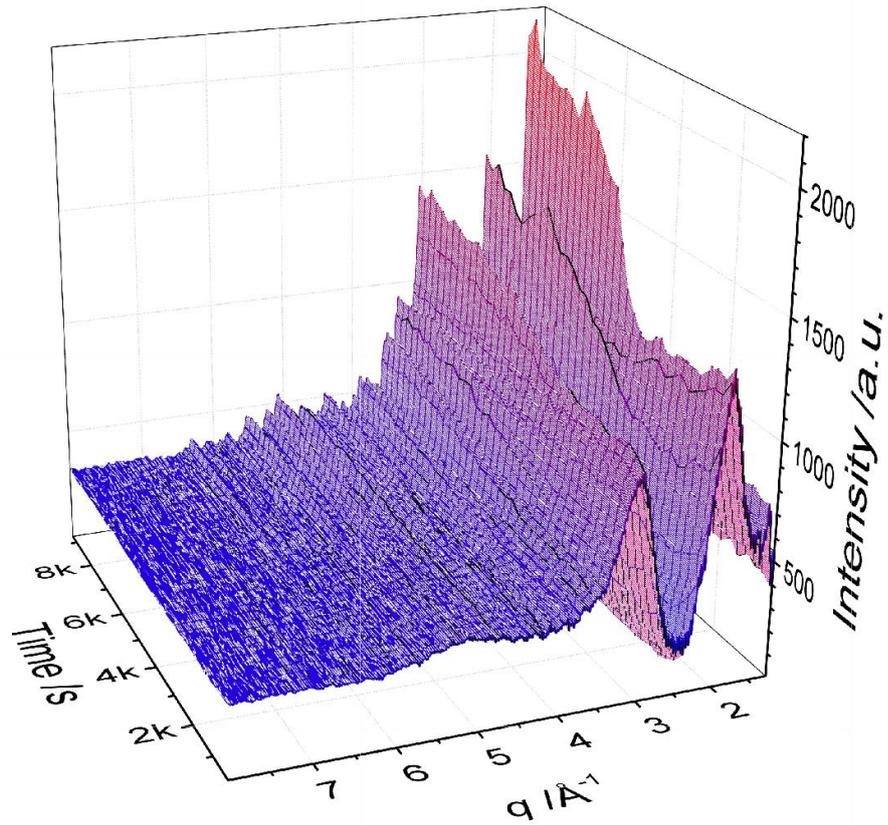

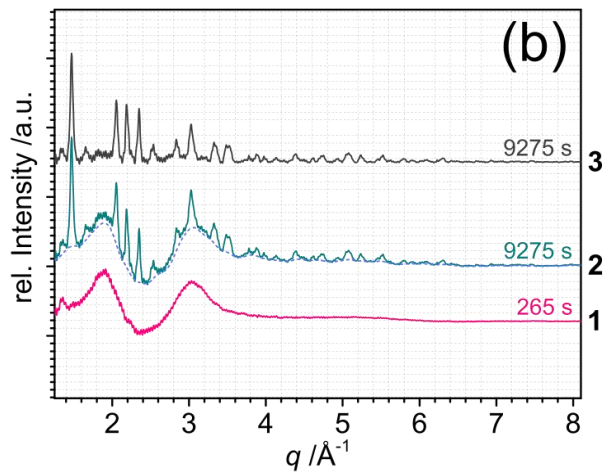
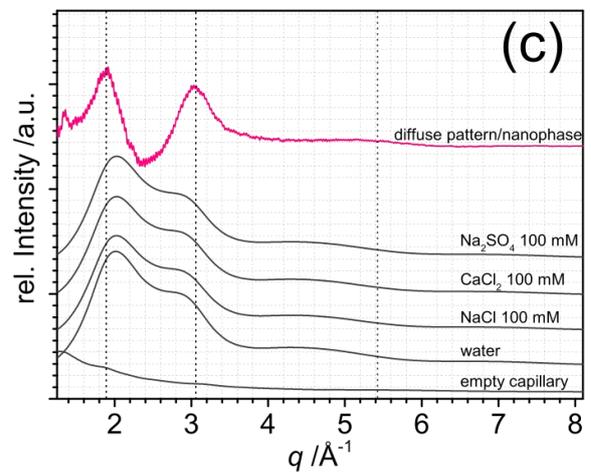
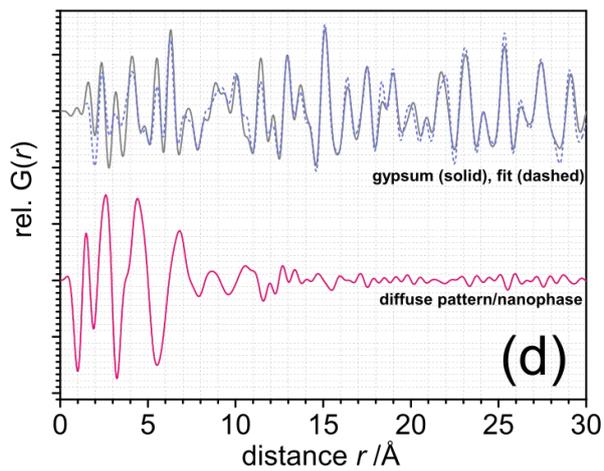
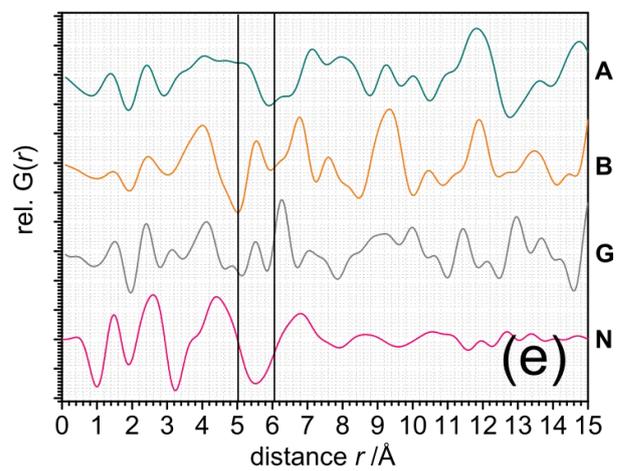

**Fig. 1**. Experimental data that follow the formation of solids in a solution with an initial concentration of 50 mM $CaSO_4$ at 10 °C. (a) time-resolved in situ HEXD pattern in which the first frame corresponds to 265 s and the last to 9,275 s; (b) comparison between the diffraction patterns of the initial nanophase 1 (solid, pink) and the final crystalline product 2 (solid, cyan). This final diffraction pattern is a juxtaposition of the diffraction pattern of the nanophase (dashed, blue) and the second purely crystalline phase 3 (solid, black); (c) comparison between the diffraction patterns of the nanophase and other components of the investigated system; (d) selected PDFs from (a) highlighting the structural difference between the nanophase and the crystalline phase clearly identified as gypsum; (e) comparison of PDFs from the nanophase - N, and other $CaSO_4$ phases: gypsum - G, bassanite - B, and anhydrite - A. The difference in the $G(r)$s for the distance $r$ ~5-6 Å is highlighted with two black lines. For (b)-(e) offset is introduced for clarity.

The four plausible clusters (Fig. 2; see ESI for the structural files) were optimized based on the following cross-sectional motifs: I – a simple 2-member square-shaped $Ca-SO_4-Ca-SO_4$ unit, where 6 stacks are arranged in an AB pattern and each Ca is coordinated by two in-plane water molecules; II – 3 member triangular $Ca-SO_4-Ca-SO_4-Ca-SO_4$ units, where 6 such units are stacked into a helix with a torsion of 70.4°; III - analogous to II but with a torsion of 70.7° and each Ca is coordinated by two in-plane water molecules; IV – analogous to I, but with a repetitively stacked $(Ca-SO_4)_2$ unit for which the initial positions were taken from a gypsum fragment (which results in a longer length of the cluster). We used a Pearson correlation coefficient ($R$) as a proxy to benchmark the structures based on their simulated $G(r)$s with respect to the experimental one. All the proposed models reproduce almost ideally the experimental $G(r)$ up to $r = 6$ Å ($R > 0.99$), with minor deviations for higher $r < 12$ Å. Our analysis shows that the proposed family of solutions should fulfill three conditions: (1) the clusters are rod-shaped; (2) a single structural unit in the cross-section of the rod does not expand beyond a single Ca-Ca distance, with either of 2 or 3 Ca ions per unit. The diameter of such a cross-sectional unit is ~0.5 nm; (3) the rods are "internally" an anhydrous $CaSO_4$ structure, and bound water is only present at their "surface"; the presence of the coordinated water improves measurably the quality of the fits for $r > 6$ Å (i.e. compare otherwise identical clusters II and III with $R$ of 0.868 vs. 0.960).

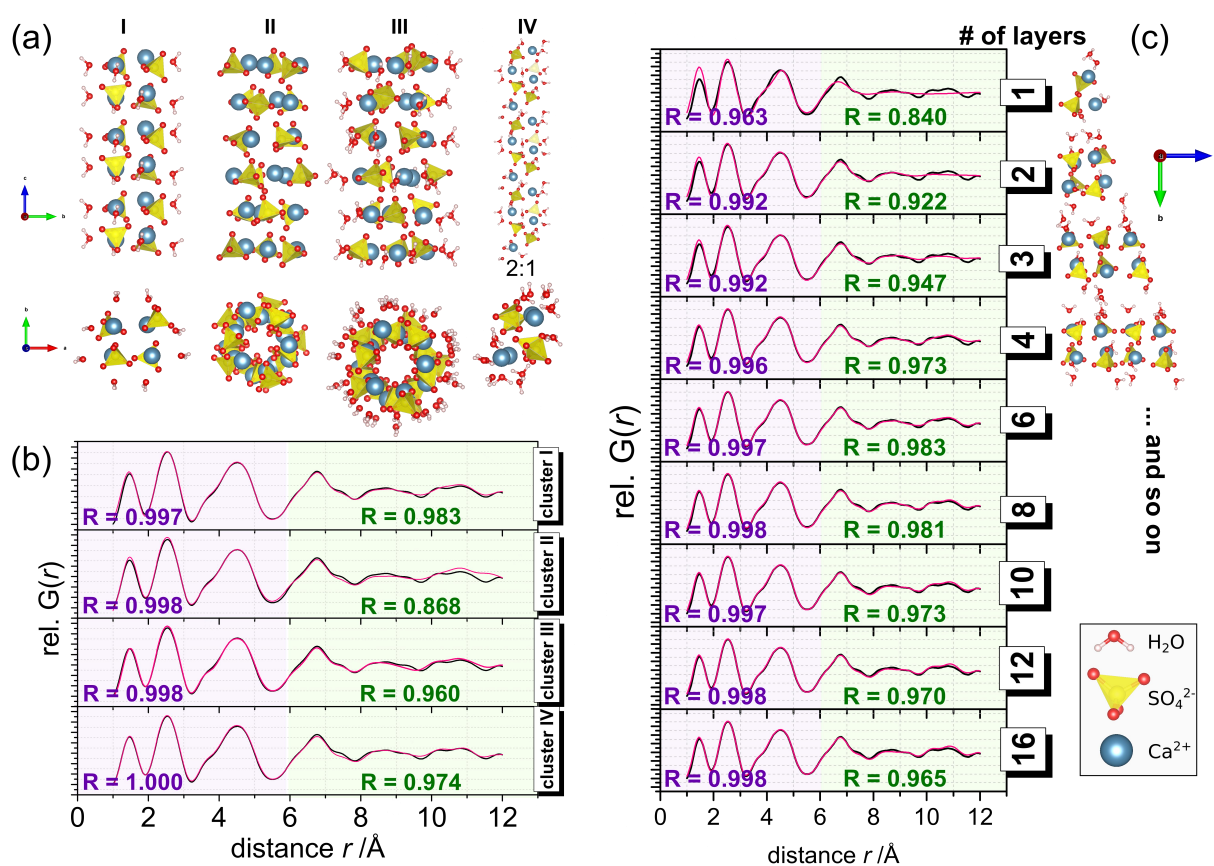

**Fig. 2.** (a) Selected optimized structural models of $CaSO_4$ rod-shaped clusters for the observed nanophase. Upper row: longitudinal view, Lower row: cross-sectional view. Cluster: I – Ca-Ca AB stack hydrated with 6 layers, II – Ca-Ca-Ca helix stack 70.4° anhydrous with 6 layers, III Ca-Ca-Ca helix stack 70.7° hydrated with 6 layers, IV – Ca-Ca AA stack hydrated with 6 layers. See ESI for the files containing these models; (b) Experimental PDF from the nanophase compared with the PDFs simulated from the structural models of clusters

I-IV; (c) Experimental PDF vs. simulated PDF for structures based on cluster III, in which the number of layers was varied from 2 to 30.

In Fig. 2C we compare the simulated PDFs of clusters similar to type I (with 6 layers) with the experimental data varying the number of layers from 1 to 16. All the clusters considered reproduce the experimental $G(r)$ up to $r$ = ~6 Å (R > 0.99). However, the profile of the PDF is reproduced better for longer clusters with 4 to 16 layers, which is also reflected by the higher $R$-values in comparison with clusters containing 1, 2 or 3-layers for $r$ > ~6 Å. As the number of layers is further increased the $R$-values for $r$ > ~6 Å start to marginally decrease again, but they remain relatively high at $R$ > 0.9. The values of $R$ suggest that the optimum structure has 4 to 10 layers (length between 1.1 and 3.1 nm), but the clusters could be longer. Nonetheless, taking into account the overall coherence length, which was limited to ~1.5 nm (Fig. 1E), we can calculate the maximum physical length of a cluster. For anisotropic rod-shaped species, we can use their radius of gyration, $R_g$, as a proxy for the measured isotropic coherence length by using the dimensions of a cylinder of length $L_c$ and radius $R_c$, and the following equation[25] $R_g^2 = R_c^2/2 + L_c^2/12$. The clusters considered have a cross-sectional diameter[12] of ~0.50 nm, which results in a length that does not exceed 5.2 nm. This in turn is equivalent to a cluster composed of 16 layers at most. Similar trends were found for structures I, III and IV, where the optimal fits are obtained for rods composed of ~4-12 stacks of structural units, yielding a total length for the rod in the range of ~1.5 - 3.5 nm depending on the packing of the structural units. The estimate of the length of the clusters obtained from the PDF analysis is in accordance with the earlier SAXS measurements of these species[12,13].

Not only do diffraction data allow us to resolve the possible structures of the observed clusters. They are also useful to evaluate the population of such species with respect to the initial ion concentrations and the final expected gypsum content by calculating the scattering invariants of the crystalline and disordered phases. We performed such analysis using diffraction data of gypsum precipitating at 10 ºC from solution containing initially 50 and 100 mM $CaSO_4$. For both conditions, diffraction patterns of an identical nanophase profile were collected, but as expected, the crystallization process progressed faster for the higher initial $CaSO_4$ concentration[13]. As shown in Fig. 3, the nanophase constituted ~100% of the initial scattering invariant for the 50 mM solution, and >90% for the 100 mM $CaSO_4$. In both cases, by the end of the experiment the nanophase was still persistent and constituted in fact the majority contribution to the invariant at ~82% and ~75% for 50 mM and 100 mM $CaSO_4$ solutions, respectively. Based solely on the HEXD data, it is not possible to determine whether the diffuse pattern originated from solute clusters or from clusters aggregated into larger morphologies. Using thermodynamic modelling approach, we evaluated the expected initial speciation, the final composition of the solution at the equilibrium, and the amount of the precipitated solid for the $CaSO_4$ solution under the as-defined physicochemical conditions (Table 1). Such an approach allowed us to calculate the maximum yield of gypsum (the precipitated solid) that one could obtain from a solution of a given starting concentration of ions (but only including the phases which are in the database). As we compare the final contribution of the amount of the nanophase with respect to the expected amount of gypsum from both solutions, it becomes apparent that the total population of the clusters considerably exceeds what one would expect to find within a crystalline solid, even if we assume that some the clusters could "dwell" in solution (and being equivalent to what the calculation considers to be the free ions or the ion pairs). This in turn translates into the fact that for the both considered conditions the "final" amount of crystalline gypsum is lower than anticipated at the equilibrium, and hence that the processes is not at the equilibrium by the end of the experiments (although the diffraction patterns were not evolving any more). Such an observation can be rationalized if we assume that gypsum single crystals are in fact mesocrystal-like[26] in nature, composed of crystalline domains "glued" together by the disorganized cluster ("bricks-in-the-wall")[27], and thus a large proportion of the clusters may co-exist inside the solid together with crystalline gypsum. Such imperfect crystals would mature over time converting clusters into the crystalline lattice approaching the equilibrium.

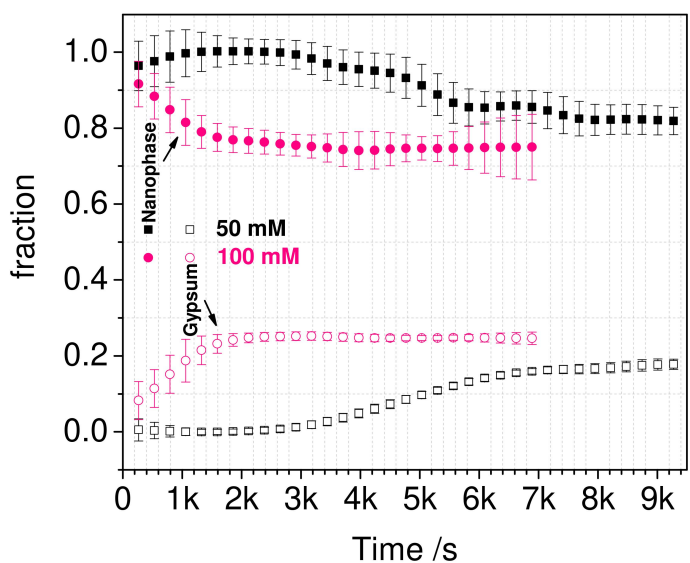

**Fig. 3.** Contribution as a function of time of the nanophase (filled symbols) and gypsum (open symbols) to the total scattering invariant with an initial concentration of 50 mM (black) and 100 mM (pink) $CaSO_4$ at 10 °C.

To complement the diffraction data, we also attempted to quantify the overall thermodynamic stability of the described above $CaSO_4$ clusters by using metadynamics[28] simulations with our recently developed[17] force field model and a suite of different collective variables. Although these simulations did not produce unequivocal values for the formation free energy of small clusters (up to 4 formula units), the aggregates did consistently show rod-like shapes (please consult our additional discussion in ESI for the further details on the issues in obtaining the reliable thermodynamic properties of the clusters from the simulations). We therefore decided to use unbiased Molecular Dynamics (MD) simulations, with rigid-ion (*i.e.* classical) and polarizable (*i.e.* AMOEBA) force fields, to assess the stability of the clusters that gave the best fit to the diffraction data (Fig. 2). The scattering-derived clusters were annealed for 5 ns at 300 K. Independent of the force field used and for all the clusters considered, only cluster I maintained its structure throughout the whole MD simulation (Fig. 4), while the remaining models significantly rearranged. The AMOEBA polarizable force field showed a high degree of instability for the remaining clusters, which tended toward elongated highly distorted structures (e.g. Fig. 4A for cluster II), while on the other hand, the rigid-ion force field seemed to favour more compact clusters that resembled two cluster I rods side-by-side. This different behaviour can be explained by considering that the rigid-ion force field gives a dissolution free energy for the anhydrous $CaSO_4$ phase[17] that is too endothermic, and therefore the ions tend to form more compact structures where they are in direct contact. A structural comparison between the relaxed geometry of cluster I at the end of the MD simulations shows that there are significant differences between the rigid-ion and polarizable force fields (Fig. 4B, upper; ESI - the trajectory files). In particular, the former model shows a much more rigid and ordered arrangement of the ions, which form a stack of squares rotated 90 degrees with respect to each other, while the AMOEBA simulations show a more mobile structure where the ions form a stack of distorted squares. Nevertheless, if we consider the overall trajectories of all the atoms in the cluster over a period of 5 ns, then despite this apparent "disordered" structure resulting from the use of the polarizable AMOEBA forcefield, this cluster structure is more consistent with the experimental data than the one from the non-polarizable model (SI: the trajectory files). These differences are evident in the Ca-Ca, Ca-S and S-S radial distribution functions, $g(r)$s (not PDFs), form trajectories obtained from the MD simulations (Fig. 4B, and ESI: trajectory files). In particular, the rigid-ion MD simulations show strong S-S peaks in the 4–5.5 Å region and a Ca – Ca peak in the 5–6 Å region (Fig. 4B, lower) that are neither in the AMOEBA MD simulations nor in the experimentally determined cluster I structure. On the other hand, the Ca-S $g(r)$ from both the rigid-ion and AMOEBA forcefield are consistent with the experimental data. Based on the above results, the AMOEBA forcefield seems to outperform the non-polarizable rigid ion forcefield and cluster I is the most likely candidate for the model structure of the $CaSO_4$ clusters as it reproduces the recorded scattering pattern. From the MD simulations, it is evident that the terminal parts of the cluster are more mobile, due to having only half of the ionic interactions, which suggests that the core part of the rod is contributing the most to the scattering signal, and therefore the clusters might be slightly longer than the 4-12 stack length that best fits the scattering data. Furthermore, a highly dynamic character is intuitively expected for such small species in solution, yet is practically impossible to implement in our fitting routines used to obtain the rigid structural models in Fig. 2 from the diffraction data. The imposition of constraints results from the fact that we had to limit the number of degrees of freedom to achieve convergence of the template models.

Typically, such constraints are valid assumptions for crystalline structures and are based on symmetry restrictions. Nonetheless, for small clusters such constraints yield merely a valid transitory averaged model (since it fits the experimental PDF), but not necessarily the minimum energy model or any equivalent low-symmetry counterpart structure. Hence, the MD simulations are essential to probe the configuration space in which our cluster I, as defined in Fig. 2, is actually just one of many possibilities. From the MD simulations it is not possible to gauge the actual thermodynamic stability of these clusters, but dynamic models using the AMOEBA force-field strongly support the HEXD and previous SAXS data[12], which show that $CaSO_4$ forms rod-shaped structures, upon creation of a supersaturated solution, that are thermodynamically stable/kinetically persistent in solution for sufficiently long time to be observed experimentally. It is very important to also note that at present the solubility of the $CaSO_4$ phases with the AMOEBA force field is unknown as it was parametrized for the solution phase, and as such it is possible that it may underestimate the stability of the solids.

Estimations of the relative stability of a 4 formula structural unit based on cluster I relative to that of a $CaSO_4$ ion pair using *ab initio* quantum mechanical methods suggest that both species are likely to be similar in free energy (see ESI).

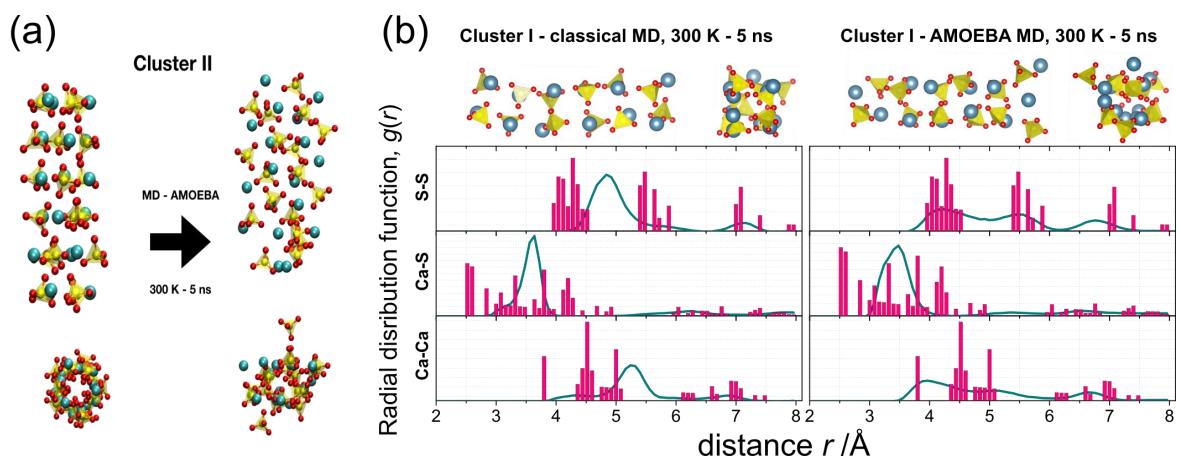

**Fig. 4.** a) Unstable cluster II after 5 ns of an MD simulation using the AMOEBA force field; the cluster shows significant disorder, but it essentially preserves its' original elongated shape; b) Comparison between the structures of cluster I as predicted from the rigid-ion classical (left) and AMOEBA polarizable (right) MD simulations. The models show the structures after 5 ns. The graphs show the radial distribution functions ($g(r)$s) computed from the whole MD trajectories (see ESI for the files) over the course of 5 ns (cyan lines) overlaid with those computed from the static structure of cluster I obtained from the experiments (magenta bars/histograms). The magenta $g(r)$ is generated from the optimized cluster I structure (see Fig. 2), hence the discrete histogram form.

**Implications**

Although our current diffraction and earlier scattering data[12,13], as well as the current MD simulations, clearly point to the presence of small $CaSO_4$ clusters as a precursor to gypsum and possibly other calcium sulfate crystalline phases[29], it is not clear what their role exactly is. It appears that they might be akin to prenucleation clusters reported for the calcium carbonate system[4,11], though our experimental approach considers in fact supersaturation conditions.

Regardless of the classification of the clusters (e.g. PNCs), their presence and evolution imply that the early stages of gypsum formation from solution are non-classical[1]. Important is also the fact that these primary clusters are numerous and persistent both in aqueous solution as well as later in the solid. Their presence along the crystalline gypsum is consistent with the surface structure factor theory of surface fractal aggregates that we developed recently[27], and the associated SAXS results which showed that the large aggregates of clusters had formed before any crystalline phases were present and persisted after gypsum crystallization[12]. Most significantly, we recently confirmed using dark field transmission electron microscopy the actual "brick-in-the-wall" (mesocrystalline) character of various natural and synthetic single $CaSO_4$ crystals[30]. These earlier results, together with the current data, strongly support our conclusion that gypsum crystallizes from the inside of large aggregates of structurally distinct, but flexible, precursor clusters.

**Conclusions**

In this work we used in situ high-energy X-ray diffraction experiments and molecular dynamics simulations to derive the structure of the precursor clusters of crystalline gypsum. We fitted several plausible cluster structures to the derived pair distribution functions (PDFs), and show that the proposed family of solutions has to fulfill three conditions: (1) the clusters are rod-shaped; (2) a single structural unit in the cross-section of the rod does not expand beyond a single Ca-Ca distance, with either of 2 or 3 Ca ions per unit; (3) the cluster structure are "internally" anhydrous and bound water is only present at their "surface". The dynamic properties of the plausible structures derived from PDF analysis were tested using unbiased MD based on both rigid-ion and polarizable force-fields. Of the suite of clusters tested, the one with a simple 2-member square-shaped Ca-SO$_4$-Ca-SO$_4$ unit, where 6 stacks are arranged in an AB pattern and each Ca is coordinated by two in-plane water molecules, resulted to be the only structure stable throughout the whole MD simulation (5 ns). Although it was not possible to gauge the actual thermodynamic stability of these clusters, the dynamic models using the AMOEBA force-field strongly support the HEXD and previous SAXS data[12], showing that CaSO$_4$ forms rod-shaped structures, upon creation of a supersaturated solution, are thermodynamically stable/kinetically persistent in solution and play a crucial role in the formation pathway of gypsum.

**Conflicts of interest**

There are no conflicts to declare.

**Acknowledgements**

This research was made possible by a Marie Curie grant from the European Commission: the NanoSiAl Individual Fellowship, Project No. 703015 to TMS. We also acknowledge the financial support of the Helmholtz Recruiting Initiative grant No. I-044-16-01 to LGB. The authors would like to thank Diamond Light Source for beamtime (proposal EE11320), and the staff of beamlines I15, in particular Dr. Philip Chater for assistance with HEXD data collection. AESVD acknowledges financial supported by the French national programme EC2CO - Biohefect, SULFCYCLE. P.R. and J.D.G. thank the Australian Research Council for funding through FT130100463 and FL180100087. The Australian National Computational Infrastructure (NCI) and the Pawsey Supercomputing Centre are acknowledged for the provision of computing resources through the NCMAS and Partners allocation schemes. E.H.B thanks the Department of Chemistry at Curtin for her PhD scholarship.

**Electronic Supporting Information description**

High-energy X-ray diffraction measurements and data analysis (**INCLUDED with this pre-print**)

**Available at:**

**https://github.com/tomaszstawski/CaSO4-Nanorods-Supporting-Information**

a) Ab initio calculation of cluster stability.

b) Structural files of Ca-SO$_4$ clusters from Fig. 2 in the main text are included as *cluster_I.cif, cluster_II.cif, cluster_III.cif* and *cluster_IV.cif*.

c) Trajectory files from the MD simulations for cluster I from which Fig. 4B in the main text was derived are included as *cluster_I_classical_trj.xyz* and *cluster_I_amoeba_trj.xyz*.

*Electronic Supporting Information*

*to*

**The Structure of CaSO$_4$ Nanorods - the Precursor of Gypsum**

*Tomasz M. Stawski[1]\*, Alexander E.S. Van Driessche[2]\*\*, Rogier Besselink[1,2],*

*Emily H. Byrne[3], Paolo Raiteri[3]\*\*\*, Julian D. Gale[3], Liane G. Benning[1,4,5]*

***Supporting Discussion***

*High-energy X-ray diffraction measurements and data analysis*

High-energy X-ray diffraction (HEXD) measurements were carried out at beamline I15 (Diamond Light Source (UK), at a wavelength of 0.16416 Å allowing us to collect scattering in the *q*-range up to ~25 Å$^{-1}$. The q-range was calibrated using a CeO$_2$ standard in a 1 mm Kapton capillary. The 2D HEXD patterns were collected using a Perkin Elmer flat panel 1621 EN detector (2048 x 2048 pixels, pixel size 0.2 x 0.2 mm$^2$) in Debye-Scherrer geometry (according to the I15 guidelines) at a time-resolution of 30 s/frame for a series of 5 data frames, after which two detector dark current images (the 6th and the 7th frame) were collected for 30 s. Including the instrumental dead-time, each such series lasted 265 s. Dark current contributions were corrected automatically by the acquisition software. Hence, in our experiments, the time resolution was thus effectively limited to 265 s because the collected frames had to be bundled and averaged to optimize the signal-to-noise, signal-to-background ratios and corrected for the dark current of the CCD detector. Furthermore, various background and reference samples were measured under similar conditions, and included among others, an empty Kapton capillary, a capillary containing pure water and the constituent solutions at T = 10 °C, as well as, powders of solid $CaSO_4$ phases. The as-obtained 2D images were reduced to raw 1D curves using DAWN[1,2]. Further processing and correction of raw 1D data curves involved subtraction of background solutions and extraction of characteristic patterns. Due to excluded volume effects, the contributions of solvent to the scattering patterns were smaller than unity and not constant over time. To correct for this effect, we filtered out sharp gypsum peaks in the time resolved data sets, by calculating the minimum over moving window functions and smoothing the resulting set of minima. Then, we fitted a linear combination of background patterns from capillaries with water, constituent solutions and an empty capillary to the peak-filtered data sets. From this approach, we obtained negative contributions of several solutions, which clearly indicated overfitting of the data sets with almost identical patterns of water and the constituent aqueous solutions. This showed us that we could reduce the number of patterns used for background subtraction down to two essential patterns: a capillary with an NaCl solution and an empty capillary. After performing the background subtraction with this reduced set, we observed a decrease of the contribution of the background coefficient attributed to a capillary with a NaCl solution from 0.85 to 0.75. This in turn suggested an increase in the excluded volume by 10%. Following this approach, we obtained reliable background correction factors and used it to subtract the water/capillary background from all the time-resolved data sets (including those without gypsum diffraction peaks). The diffuse and crystalline scattering patterns were separated from each other by the abovementioned moving minimum window function and the concentration profiles were derived by applying a linear combination fitting method with the derived diffuse and crystalline patterns. After this processing and background subtraction stage the atomic pair distribution functions (PDFs) were obtained by using the PDFgetX3 software package[3]. The background and Compton scattering subtractions were also performed using the same code. Further processing of the data, such as phase identification and fitting of the structural models, principal component analysis (PCA) on the series etc. was performed using custom scripts developed by us in NumPy (Python) and utilizing a dedicated DiffPy library[18,19]. To achieve convergence of the results we introduced several reasonable structural restrictions: (1) the sulfate tetrahedra and water molecules were considered to be internally stiff with a free rotation around Euler angles; (2) model clusters were defined by a network of linked bond-vectors and bond distances were restricted within ±1 Å from the initial average values known from crystalline $CaSO_4$ phases; (3) to avoid that some atoms get unrealistically close or far from each other we applied a simple potential $V = \sum (k_e \cdot q_i \cdot q_j \cdot r_{ij}^{-1} + k_{i,j,\,rep} \cdot r_{ij}^{-12})$, where $q_i$ and $q_j$ are the oxidation number of atoms i and j, $r_{ij}$ is the distance between atoms i and j, $k_e$ is a Coulomb constant, and $k_{i,j,\,rep}$ is a repulsive constant optimized to known atom distances.

*Ab initio calculation of cluster stability*

Because of the dynamic nature of the clusters, as observed in the atomistic simulation, it is difficult to obtain an accurate estimate of the free energy of species via static quantum mechanical methods. In the present study, we have performed a number of cluster calculations to estimate the stability of CaSO$_4$ ion pair and a four formula unit cluster, with an approximately cubic structure, that was initially extracted from the AMOEBA molecular dynamics with its' first solvation shell. All structures were fully optimized at the B97X-D3/def2-TZVP level of theory using the ORCA 4.1.1 code[6] with Grid4 for the numerical integration of the density. In order to describe the aqueous environment of the species the ion pair and cube cluster were surrounded by 18 and 30 waters of solvation, respectively, in addition to the use of the CPCM continuum solvation model (the radius for Ca was refitted to 1.824 Å to reflect the hydration of the cation, rather than the neutral atom). The aim was to try to ensure that each ion has at least a complete first solvation shell of explicit water. Indeed, during optimization of the ion pair two water molecules relaxed from an initial position in the first solvation shell to become part of the second shell, thus confirming the completion of the first shell. For the cube, all ions remained in the first shell. Given that all Ca ions had three waters coordinated, while the sulfate ions had a mix of 4 and 5 waters of hydration (in addition to bridging waters from Ca), it is possible that extra water could have been accommodated.

Based on the above calculations, the difference in internal energy between the CaSO$_4$ ion pair and (CaSO$_4$)$_4$ cluster was computed to be 55.0 kJ/mol, with the ion pair being more stable. Direct use of the vibrational frequencies to estimate the free energy difference was found to lead to poor results as the energetics are dominated by the zero-point energy of the water molecules surrounding the species. As a result, the free energies were sensitive to the solvation of the water, rather than the CaSO$_4$, which given the single static structure is not representative of the liquid phase. Instead, it is possible to estimate the free energy difference based on the change in the number of water molecules in the first solvation shell of the ions between the clusters. From the work of Dunitz[7] each molecule of water released during aggregation could contribute up to 29.3 J/K/mol to the entropy of the system, which we have previous used to rationalize the stability of ion pairs in other systems[8]. For the ion pair and cube the optimal first shell water coordination numbers are estimated to be 16 and 32, respectively, which leads to an upper bound to the entropic stabilization of the cube of 69.9 kJ/mol relative to the ion pair. Given the similar magnitude, but opposing signs, of the internal energy and entropy for aggregation of the ion pair to the smallest motif for cluster I we can conclude that formation of such larger structures is feasible, though within the uncertainties of the method it is not currently possible to give an accurate quantitative free energy for the process.